\documentstyle[12pt,aasms4] {article}
\newcommand{\beq}{\begin{equation}}
\newcommand{\eeq}{\end{equation}}
\newcommand{\beqn}{\begin{eqnarray}}
\newcommand{\eeqn}{\end{eqnarray}}

\newcommand{\pa}{\partial}
\newcommand{\no}{\noindent}

\newcommand{\Od}{{\rm O}}

\newcommand{\non}{\nonumber}

\begin{document}
 
\title{\bf{ Bifurcation for Dynamical Systems of Planet-Belt Interaction} }
\author{Ing-Guey Jiang$^{1}$  and Li-Chin Yeh$^{2}$}

\affil{{$^{1}$ Institute of Astronomy,}\\
{ National Central University, Chung-Li, 
Taiwan} \\  
\ \\
{$^{2}$ Department of Mathematics,}\\
{ National Hsinchu Teachers College, Hsin-Chu, Taiwan}}

\authoremail{jiang@astro.ncu.edu.tw}

\begin{abstract}
The dynamical systems of planet-belt interaction are studied by the 
fixed-point analysis and the bifurcation of solutions on the parameter
space is discussed. 
For most cases, our analytical and numerical results 
show that the locations of fixed points are 
determined by the 
parameters and these fixed points are either structurally
stable or unstable.  
In addition to that, there are two special fixed points:
the one on the inner edge of the belt is asymptotically stable and the one
 on the outer edge of the belt is unstable. This is consistent with the 
observational picture of Asteroid Belt between the Mars and Jupiter: 
the Mars is moving stablely close to the inner edge but the Jupiter is
quite far from the outer edge.
\end{abstract}

\newpage

\section{Introduction}

The discovered number of extra-solar planets is increasing dramatically  
due to astronomers' observational effort, therefore the dynamical study
in this field is getting important. Because the belts of planetesimals often 
exist among planets within a planetary system as we have in the Solar System, 
it is indeed important to understand the solutions of dynamical systems of 
planet-belt interaction. Jiang \& Ip [2001] predicted that the interaction 
with the belt or disc might bring the planetary system of upsilon Andromedae
to the current orbital configuration. 

Yeh \& Jiang [2001] used phase-plane analysis to study the orbital 
migration problem of scattered planets. They completely classify the 
parameter space and solutions
and conclude that the eccentricity always increases if the planet, which 
moves on circular orbit initially, is scattered to migrate outward.
These analytical results is consistent with the numerical simulations
in Thommes, Duncan \& Levison [1999].

In addition to astronomy, general or Newton's dynamical systems 
are studied in many other fields and have very important applications. 
Clausen et al. [1998] studied periodic modes of motion of a few 
body system of magnetic holes both experimentally and numerically.
Kaulakys et al. [1999] showed that a systems of many bodies
moving with friction can experience a transition to chaotic behavior. 

On the other hand,  Chan et al. [2001] studied bifurcation for limit cycles 
of quadratic systems interestingly. Similar type of approach should be also 
good for the bifurcation of solutions for dynamical systems of 
planet-belt interaction. In this paper, we focus on the planet-belt 
interaction and study the bifurcation of such system by phase plane analysis.

Basicly, we would like to understand the orbital evolution of 
a planet which moves around a central star and  interacts with a belt.
The belt is a annulus with inner radius $r_1$ and
outer radius $r_2$, where  $r_1$ and $r_2$ are assumed to be 
constants. We set $r_1=3$ and $r_2=6$ for all numerical results 
in this paper. 

We assume the distance between the central star and the planet is $r$,
where $r$ is a function of time. 
When $r < r_1$, the belt would only give the force 
which pulls the planet away from the central star.
When $r > r_2$, the belt would only give the force 
which pushes the planet towards the central star. When $r_1 \le  r \le r_2$,
in addition to the usual gravitational force, there is friction 
between the planet and the belt. 
These three cases will be studied in Model A ($r<r_1$), Model B($r>r_2$) and 
Model C($r_1<r<r_2$) individually.

We will mention our basic governing equations in Section 2.
 In Section 3-5, we study the locations and stabilities of fixed points 
by phase plane analysis for Model A, B and C. 
The conclusions will be in Section 6.

\section{The Model}

In general, the equation of motion of the planet is (Goldstein 1980)
\beq
\frac{d^2u}{d\theta^2}=-u-\frac{mf\left(\frac{1}{u}\right)}{l^2u^2},
\label{eq:1}
\eeq
where $u=1/r$, $m$, $l$ are the mass and the angular momentum of the planet and
$f$ is the total force acting on the planet.
We use the polar coordinate $(r,\theta)$ to describe the 
location of the planet.

The total force $f$ includes the contribution from the central star 
and the belt. The force from the central star is 
\beq
f_s=-\frac{Gm}{r^2},\label{eq:fs}
\eeq
where we have set the mass of the central star to be one.
 If  
the density profile of belt is 
$\Sigma(r)={c_0}/{r}$ where $c_0$ is a constant completely
determined by the total mass of the belt. 
The total mass of the belt is 
\beq
M_{b0}=\int^{2\pi}_{0}\int^{r_2}_{r_1}\Sigma(r')r'dr'd\phi =2\pi c_0(r_2-r_1) 
\eeq

In general, the force from the belt for the planet is complicated and involved
the Elliptic Integral.  We use a simpler integral to over-estimate the 
force and then use a small number $0<\beta<1$ to correct the force 
approximately. 

When the planet is not within the belt region, the force from the belt is 

\beqn
f_{b\pm} &=& \pm \beta\int^{\pi}_{-\pi}\int^{r_2}_{r_1} \frac{Gm\Sigma(r')r'}
{r^2+(r')^2-2rr'\cos\phi} dr'd\phi \non \\
&=&\pm\frac{\beta\pi Gc_{0}m}{r}\left\{\ln\biggm|\frac{(r_1-r)(r+r_2)}
{(r_2-r)(r+r_1)}\biggm|\right\} \non \\
&=&\pm\frac{\pi Gcm}{r}\left\{\ln\biggm|\frac{(r_1-r)(r+r_2)}
{(r_2-r)(r+r_1)}\biggm|\right\},\label{eq:fd}
\eeqn
where $+$ stands for the case when $r < r_1$ (the belt pulls the planet
away from the central star), 
$-$ stands for the case when $r > r_2$
(the belt pushes the planet towards the central star) and we define 
$c=\beta c_{0}$.

When the planet is within the belt region, the gravitational force
from the belt is 
\beqn
f_{b\epsilon}&=&-\beta\int^{\pi}_{-\pi}\int^{r-\epsilon}_{r_1} 
\frac{Gm\Sigma(r')r'}{r^2+(r')^2-2rr'\cos\phi} dr'd\phi 
+\beta\int^{\pi}_{-\pi}\int^{r_2}_{r+\epsilon} \frac{Gm\Sigma(r')r'}
{r^2+(r')^2-2rr'\cos\phi} dr'd\phi, \non \\
&=&\frac{\pi Gcm}{r}\ln\left\{\biggm|\frac{(r+r_1)(r+r_2)}
{(r-r_1)(r-r_2)}\biggm|\left(\frac{\epsilon^2}{4r^2-\epsilon^2}\right)
\right\},
\eeqn
where $\epsilon$ is a small number to avoid the singularity of the integral.
To be convenient, we define the effective mass of the belt to give force on 
the planet to be 
\beq
M_{b}\equiv \beta M_{b0}=2\pi c(r_2-r_1). 
\eeq

Because there might be some scattering between the planetesimals in the  
belt and the planet,
we should include frictional force in this case. This frictional force 
should be proportional to the surface density of the belt at the location
of the planet and the velocity of the planet. However, if the planet is doing 
circular motion, the probability of close encounter between the planet and
the planetesimals in the belt is very small and can be neglect here. Thus,
we can assume that the frictional force is proportional to the radial
velocity of the planet  $dr/dt$ only and ignore the $d\theta/dt$  dependence.

Therefore,  $f_{\alpha}$ is proportional to the surface density of the 
belt and radial velocity $dr/dt$. Hence, we write down the formula for 
frictional force as

\beq
f_{\alpha}=-\alpha \Sigma(r)\frac{dr}{dt}= -\frac{\alpha c}{r}\frac{dr}{dt},
\label{eq:force1}
\eeq
where $\alpha$ is a frictional parameter. 
Because the  $d\theta/dt$  component of the planetary velocity 
is ignored in the frictional force,  
the angular momentum $l$ is conserved here, so we have 
\beq
mr^2{d\theta}=l{dt}. \label{eq:angu_mom}
\eeq
Because of this, we can use $\theta$ as our independent variable.
We use $\theta$ to label time $t$ afterward and one can easily
gets $t$ from the above equation. 

Since $u=1/r$ and equation (\ref{eq:angu_mom}), we have 
\beq
\frac{dr}{dt}=\frac{dr}{d\theta}\frac{d\theta}{dt}
=\frac{l}{mr^2}\frac{dr}{d\theta}=-\frac{l}{m}\frac{du}{d\theta}.
\label{eq:drdt}
\eeq

Equation (\ref{eq:force1}) becomes
\beq
f_{\alpha}=\frac{l \alpha cu}{m}\frac{du}{d\theta}.\label{eq:force3}
\eeq

\section{Model A}

In this model, we assume that the star-planet distance is less than 
the inner radius of the belt, i.e. $r < r_1$. Thus,
$f=f_s+f_{b+}$, where $f_s$ and $f_{b+}$ are defined in Section 2.
Since $r<r_1<r_2$, from equation (\ref{eq:fd}),
we have

$$f_{b+}=\frac{\pi Gcm}{r}\left\{\ln\biggm|\frac{(r_1-r)(r+r_2)}
{(r_2-r)(r+r_1)}\biggm| \right\}.$$

We further define that  $u\equiv 1/r$ and also
\beq
P_I(u)\equiv 
\biggm|\frac{(r_1u-1)(1+r_2u)}{(r_2u-1)(1+r_1u)}\biggm|. \label{eq:p_i}
\eeq

We can then transform equation (\ref{eq:1}) into 
\beq
\frac{d^2u}{d \theta^2}=-u+\frac{Gm^2}{l^2}-\frac{\pi Gcm^2}{l^2u}
\left\{\ln P_I(u) \right\}. \label{eq:p1I1}
\eeq

Therefore, the equation of motion for this system can be written as 

\begin{eqnarray}
 \frac{du}{d\theta}&=&v \\
\frac{dv}{d\theta}&=&-u+k_2-\frac{k_3}{u}\ln (P_I(u)),
\end{eqnarray}\label{eq:p1I2}

where  $k_2=Gm^2/(l^2)$ and $k_3=(c\pi Gm^2)/(l^2)$.

\subsection{Fixed Points for Model A}

The fixed points $(u,v)$ of problem (\ref{eq:p1I2}) satisfy the following 
equations

$$v=0$$
\beq
{\rm and}\qquad -u+k_2-\frac{k_3}{u}\ln(P_I(u))=0.\label{eq:p1I3}
\eeq

Obviously, only those solutions locating in the region that
$u > 1/r_1$ have physical meaning since we consider the case that 
$r < r_1$ is this model. However, the properties of the fixed points
in unphysical region would affect the topology of the solution curves
in physical region, so we have to study fixed points both in physical
and unphysical regions.
 
If we define  
\beq 
Q_A(u)=\exp\left\{-\frac{u^2}{k_3}+\frac{k_2}{k_3}u \right\},
\label{eq:model_a_qa}
\eeq
from equation (\ref{eq:p1I3}), we know that 
fixed points should satisfy $P_I(u)=Q_A(u)$. 
In the following theorem, we prove that there are at least 
three fixed points in
whole phase space and at least one locates within the physical region,
$u>1/r_1$.

In our theorems, we denote $(1/r_1)^{+}$ to 
represent that $u$ tends to $1/r_1$ from the right hand side 
and $(1/r_1)^{-}$ to 
represent that   
$u$ tends to $1/r_1$ from the left hand side etc.

{\bf Theorem 3.1}

(a) There is at least one $u_{\ast}<1/r_2$ such that  
$P_I(u_{\ast})=Q_A(u_{\ast})$.

(b) There is at least one $u_{\ast\ast}\in (1/r_2,1/r_1)$ such that 
$P_I(u_{\ast\ast})=Q_A(u_{\ast\ast})$.

(c) There is at least one $u_{\ast\ast\ast}>1/r_1$ such that 
$P_I(u_{\ast\ast\ast})=Q_A(u_{\ast\ast\ast})$.

{\bf Proof :} 

Let $R_A(u)\equiv P_I(u)-Q_A(u)$,
where $P_I(u)$ and $Q_A(u)$ are defined in equation (\ref{eq:p_i}) and 
equation (\ref{eq:model_a_qa}).

(a) Since 
$$R'_A(u) =\frac{2(r_2-r_1)(1+r_1r_2u^2)}{[1-(r_2-r_1)u-r_1r_2u^2]^2}
-\exp\left\{-\frac{u^2}{k_3}+\frac{k_2}{k_3}u \right\}
\left\{-\frac{2u}{k_3}+\frac{k_2}{k_3}\right\},$$
we have
\beq 
R'_A(0)=2(r_2-r_1)-\frac{k_2}{k_3}=2(r_2-r_1)-\frac{1}{\pi c}
=2(r_2-r_1)-\frac{2(r_2-r_1)}{M_b}
=2(r_2-r_1)\left(1-\frac{1}{M_b}\right).
\eeq
Because the total mass of the belt is assumed to be less than the mass of 
the central star, i.e. $0<M_b<1$, 
we have $R'_A(0)<0$. Since $R_A(0)=0$, $R'_A(0)<0$, and  
$R_A\left((1/r_2)^{-}\right)>0$, we have that there exists 
$u_{\ast}\in (0,{1}/{r_2})$ such that  $R_A(u_{\ast})=0$, that is, 
$P_I(u_{\ast})=Q_A(u_{\ast}).\Box$

(b) Since $R_A\left((1/r_2)^{+}\right)>0$ and 
$R_A((1/r_1)^{-})<0$, we have that there exists $u_{\ast\ast}\in (1/r_2,1/r_1)$
 such that  $R_A(u_{\ast\ast})=0$, that is, 
$P_I(u_{\ast\ast})=Q_A(u_{\ast\ast}).\Box$

(c)  In the region of $u>1/r_1$, we have 
$R_A(1/r_1)=P_I(1/r_1)-\exp\left\{-\frac{r_1^2}{k_3}+\frac{k_2}
{k_3}r_1\right\}<0$.

Further, because
$\exp\left\{-\frac{u^2}{k_3}+\frac{k_2}{k_3}u\right\} \to 0 $ as 
$u\to \infty$, $P_I(u)\to 1$ as $u\to \infty$ and 
$R_A(u)\to 1$ as $u\to \infty$, there is a $u_{\ast\ast\ast}> 1/r_1$ 
such that $R_A(u_{\ast\ast\ast})=0$, i.e. 
$P_I(u_{\ast\ast\ast})=Q_A(u_{\ast\ast\ast}).\Box$

Figure 1(a)-(b) are the numerical results for the solutions of equation 
(\ref{eq:p1I3}). That is, we find all possible fixed points 
$u$ for any given $k_2$ and $M_b$ with 
fixed values of $r_1$ and $r_2$. 
Figure 1(a) are the results on $k_2-u$ plane, where dashed lines are for
$M_b=0.1$, dotted lines are for $M_b=0.5$ and solid lines are for $M_b=1.0$.
There are three lines for each value of $M_b$ but 
only two solid lines can be seen because one solid line overlaps with 
the $k_2$-axis. Thus, there are three fixed points for any given values
of $k_2$ and $M_b$.
Figure 1(b) are the results on $M_b-u$ plane, where dashed lines are for
$k_2=0.1$, dotted lines are for $k_2=0.5$ and solid lines are for $k_2=1.0$.

From these results, we can completely determine the locations of fixed points 
for different values of $k_2$ and $M_b$, where we assume $0 < k_2 < 1$
and $0 < M_b < 1$. These results are consistent with the analytic results
of Theorem 3.1. 
\subsection{Phase-Planes for Model A}
Following the linearization analysis, 
the eigenvalues $\lambda$ corresponding to the fixed points 
$(u,v)$ satisfy the following equation

$$\lambda^2-(-1-k_3\frac{\pa A_1}{\pa u})=0,$$

where $A_1= (\ln P_I(u))/{(u)}$, so 
\beq
\frac{\pa A_1}{\pa u}=-\frac{\ln P_I(u)}{u^2}+\frac{1}{uP_I(u)}
\frac{\pa P_I(u)}{\pa u}\label{eq:model_a_a1}
\eeq


Hence 
\beq 
\lambda =\pm\sqrt{-1-k_3\frac{\pa A_1}{\pa u}}.
\eeq
 
We thus have two cases in the following: 

Case 1: If $-1-k_3\frac{\pa A_1}{\pa u}>0$, then $\lambda$ are real with 
opposite sign, so in this case, the fixed point is a unstable saddle point.

Case 2 : If $-1-k_3\frac{\pa A_1}{\pa u}<0$, then $\lambda$ are pure complex 
numbers with opposite sign, so in this case, the fixed point is a center point.

Although we understand that when the real parts of the eigenvalues of a 
fixed point equal to zero in the first-order linearization analysis, 
the properties of this fixed point should be determined by the higher
order analysis in general. However, to simply our language, we still 
use the term, center point, for any fixed point with zero real parts
of eigenvalues. This is a good choice because our numerical results
show that these points are in fact center points.

In the following theorem, we discuss the properties of the fixed point 
for the physical region, i.e. $u>1/r_1$.

{\bf Theorem 3.2 }  

If the fixed point $u^{\ast}>1/r_1$, then this 
fixed point $u^{\ast}$ is a center point.

{\bf Proof :}  

First, from equation (\ref{eq:p1I3}) and equation (\ref{eq:model_a_a1}), 
we calculate 
\beqn
-1-k_3\frac{\pa A_1}{\pa u} &= &-1+k_3\frac{\ln P_I(u)}{u^2}-
\frac{2k_3(r_2-r_1)(1+r_1r_2u^2)}{u(r_2^2u^2-1)(r_1^2u^2-1)} \non\\
&=& -1+k_3\left(-\frac{1}{k_3}+\frac{k_2}{k_3}\frac{1}{u}\right)-
\frac{2k_3(r_2-r_1)(1+r_1r_2u^2)}
{u(r_2^2u^2-1)(r_1^2u^2-1)} \non \\
&=& -2 +\frac{k_2}{u}-\frac{2k_3(r_2-r_1)(1+r_1r_2u^2)}
{u(r_2^2u^2-1)(r_1^2u^2-1)}.
\eeqn

For the convenience, we separate the proof into two parts: 

(i) when $k_2/2<1/r_1$:\\
Since $u^{\ast}>1/r_1$, we have
$u^{\ast}>k_2/2$. Thus, $-2+k_2/u^{\ast}<0$, so  
$-1-k_3\frac{\pa A_1}{\pa u}\biggm|_{u=u^{\ast}}<0$. The fixed point 
$u^{\ast}$ is therefore a center point. 

(ii) when $k_2/2>1/r_1$:\\
Since $u^{\ast}>1/r_1$ is a fixed point,  $u^{\ast}$ satisfy  
$P_I(u^{\ast})=Q_A(u^{\ast}).$
Since we consider the region $u>1/r_1$, we have 
\beq
0<P_I(u)<1 {\rm\,\, for\,\,} u>1/r_1 \label{eq:thm22_pi}
\eeq 
Moreover, because $Q_A$ only has one critical point at $u=k_2/2$, 
i.e. $Q'_A(k_2/2)=0$ and $Q''_A(k_2/2)<0$, $Q_A(k_2/2)$ is a global maximum 
and 
$$Q_A\left(\frac{k_2}{2}\right)>Q_A(k_2)=1.$$
Since $Q_A(0)=1$, $Q_A({k_2}/{2})>1$ and $Q''_A(u)<0$ for $u<k_2$,   
we have $Q_A(u)\ge 1$ for $u<{k_2}$. From equation (\ref{eq:thm22_pi}), 
we have $R_A(k_2)=P_I(k_2)-Q_A(k_2)<0$ and $\lim_{u\to \infty}R_A(u)=1$.
Therefore, we have  $u^{\ast}>k_2$. 

Because $u^{\ast}>k_2/2$, we have $-2+k_2/u^{\ast}<0$ and thus 
$-1-k_3\frac{\pa A_1}{\pa u}<0$. Therefore, the fixed point $u^{\ast}$ is a 
center point.$\Box$

The solution curves on the $u-v$ phase plane are shown in Figure 1(c)-(d),
where we set $k_2=0.2$, $M_b=0.3$. 
In Figure 1(c),  there are two vertical dotted lines, the left one 
is $u=1/r_2$ and the right one is $u=1/r_1$, which divide the $u-v$
plane into three regions. It is obvious that there is one fixed point
in each region, which is consistent with Theorem 3.1 and the fixed point
in the region of $u>1/r_1$ is a center point, which is precisely what we 
have proved in Theorem 3.2. Figure 1(d) is just the detail 
of the solution curves near this fixed point in the region of $u>1/r_1$.
These figures reconfirm the analytic results and also make the behavior of the 
solution curve clear. We thus have known the behavior of the solutions 
completely.

\section{Model B}

In this model, we assume that the star-planet distance is larger than 
the outer radius of the belt, i.e. $r > r_2$. Thus,
$f=f_s+f_{b-}$, where $f_s$ and $f_{b-}$ are defined in Section 2.
( $f_s$ is defined by equation (\ref{eq:fs})
and $f_{b-}$ is defined by equation (\ref{eq:fd}). )

In this model, since $r>r_2>r_1$, from equation (\ref{eq:fd}) in Section 2,
we have

$$f_{b-}=-\frac{\pi Gcm}{r}\left\{\ln\biggm|\frac{(r_1-r)(r+r_2)}
{(r_2-r)(r+r_1)}\biggm|\right\}.$$

By $u=1/r$ and  equation (\ref{eq:1}), we have the equation of motion: 
\beq
\frac{d^2u}{d \theta^2}=-u+\frac{Gm^2}{l^2}+\frac{\pi Gcm^2}{l^2u}
\left\{\ln \biggm| \frac{(1-r_1u)(1+r_2u)}{(1-r_2u)(1+r_1u)}\biggm|
\right\}. \label{eq:II1}
\eeq

Similarly, we set $k_2={Gm^2}/{(l^2)}$ and $k_3={(c\pi Gm^2)}/{(l^2)}$
and transform equation (\ref{eq:II1}) to the following problem:
\begin{eqnarray}
 \frac{du}{d\theta}&=&v \\
\frac{dv}{d\theta}&=&-u+k_2+\frac{k_3}{u} \ln (P_I(u)),
\end{eqnarray}\label{eq:II2}
where $P_I(u)$ is defined in equation (\ref{eq:p_i}).

\subsection{Fixed Points for Model B}
The fixed points $(u,v)$ of problem (\ref{eq:II2}) satisfy the following 
equations \\

$$v=0$$
\beq
{\rm and}\qquad -u+k_2+\frac{k_3}{u}\ln(P_{I}(u))=0.
\label{eq:II3}
\eeq

If we define  
\beq 
Q_B(u)\equiv\exp\left\{\frac{u^2}{k_3}-\frac{k_2}{k_3}u\right\},
\label{eq:q_b}
\eeq
and
\beq
R_B(u)\equiv P_I(u)-Q_B(u),
\eeq
then fixed points $(u,0)$ should satisfy $R_B(u)=0$. In the following 
theorem, we discuss the properties of fixed points in different
parameter space. 

{\bf Theorem 4.1}

For convenience, we define three regions as:

Region (I):  $u<1/r_2$ (this is the physical region for Model B),

Region (II): $1/r_2<u<1/r_1$,

Region (III): $u>1/r_1$.

(a) In Region (I),  if $k_2>1/r_2$, then there is no fixed point in 
this  $u<1/r_2$ region. 

(b) In Region (II), there is at least one fixed point. 

(c) In Region (III), if $k_2<1/r_1$, then there is no fixed point.

(d) Within Region (III), if $1/r_1<k_2/2$ and $P_I(k_2/2)>Q_B(k_2/2)$,
 then there is one unique fixed point: $u_2\in(1/r_1,k_2/2)$
and at least one fixed point $u_3 \in (k_2/2,k2)$.

(e) Within Region (III), if $1/r_1<k_2/2$ and $P_I(k_2/2)<Q_B(k_2/2)$,
then there is no fixed point in the region of $(1/r_1,k_2/2)$. 

(f) In Region (III), if $1/r_1<k_2/2$ and $P'_I(u)>Q'_B(u)$ for 
$u\in (k_2/2,k_2)$, then there is no 
fixed point.

{\bf Proof :}

 From the definition of $Q_B(u)$ in equation (\ref{eq:q_b}),
we have $Q'_B(k_2/2)=0$ and $Q''_B(u) >0$ for all $u$.
Thus,  $Q_B$ only has one critical point at 
$k_2/2$ and  $Q_B(k_2/2)$ is a global minimum. 

Moreover, because $Q_B(0)=1=Q_B(k_2)$, we have 
\beqn
&& Q_B(u)<1 {\rm \,\, for \,\,} u\in (0,k_2); \label{eq:modelb_41_qb1}\\
&& Q_B(u)>1 {\rm \,\, for \,\,} u\in (k_2,\infty).\label{eq:modelb_41_qb2}
\eeqn

(a) Since $P_I(0)=0$ and $P'_I(u)>0$, we have
\beq
P_I(u)>1 {\rm  \,\, for \,\,} u\in \left(0,\frac{1}{r_2}\right).
\label{eq:modelb_41_pi}
\eeq

Because of the condition $k_2>1/r_2$, 
from equation (\ref{eq:modelb_41_qb1}), we have  
\beq
Q_B(u)<1  {\rm  \,\, for \,\,} u\in \left(0,\frac{1}{r_2}\right).
\label{eq:modelb_41_qb}
\eeq

From equation (\ref{eq:modelb_41_pi}) and (\ref{eq:modelb_41_qb}), we have
$R_B(u)=P_I(u)-Q_B(u)>0$ for $u<1/r_2$. Therefore, there is no fixed point 
in this physical region, $u<1/r_2.\Box$

(b) Because $R_B(\left({1}/{r_2}\right)^{+})>0$ and $R_B(1/r_1)=
-\exp\{r_1^2/k_3-k_2/k_3\}<0$, there is at least one 
$u_1\in (1/r_2,1/r_1)$ such that $R_B(u_1)=0$. $\Box$

(c) Because of the condition $k_2<1/r_1$, from equation 
(\ref{eq:modelb_41_qb2}), we have
\beq
Q_B(u)>1 {\rm \,\, for \,\,} u\in \left(\frac{1}{r_1},\infty\right).
\label{eq:modelb_41_qb3}
\eeq

Since $P_I(1/r_1)=0$, $P'_I(u) >0$ and $\lim_{u\to \infty} P_I(u) =1$ , we have
\beq
0<P_I(u)<1 {\rm  \,\, for \,\,} u\in \left(\frac{1}{r_1}, \infty\right).
\label{eq:modelb_41_pi2}
\eeq

From equation (\ref{eq:modelb_41_qb3}) and (\ref{eq:modelb_41_pi2}), we have
$R_B(u)=P_I(u)-Q_B(u)<0$ for all $u\in (1/r_1,\infty)$. Therefore, there is 
no fixed point in this case. $\Box$
  
(d) In the region of $ (1/r_1,k_2/2)$, we have $R_B(1/r_1)<0$,
$R'_B(u) =P'_I(u)-Q'_B(u)>0$ and from our assumption
$R_B(k_2/2)=P_I(k_2/2)-Q_B(k_2/2)>0$, so there is one unique fixed point 
$u_2\in (1/r_1,k_2/2)$. Because $R_B(k_2)<0$ and $R_B(k_2/2)>0$,
there is at least one fixed point $u_3\in (k_2/2,k_2)$. $\Box$

(e) Because $R_B(1/r_1)<0$, $R'_B(u) =P'_I(u)-Q'_B(u)>0$ for 
$u\in (1/r_1,k_2/2)$ and from our assumption 
$R_B(k_2/2)=P_I(k_2/2)-Q_B(k_2/2)<0$, there is no fixed point in 
this $(1/r_1,k_2/2)$ region. $\Box$

 (f) First, we consider the region of $(1/r_1,k_2)$.
 Because of the condition $1/r_1<k_2/2$ and  $ R'_B(u)=P'_I(u)-Q'_B(u)>0$ for  
$u\in (k_2/2,k_2)$, from our assumption,  we have 
$$ R'_B(u)=P'_I(u)-Q'_B(u)>0 \quad {\rm for\,\, all\,\, } 
u\in \left(\frac{1}{r_1},k_2\right).$$
Further, since $R_B(1/r_1)<0$ and $R_B(k_2)<0$, there is no fixed point in 
$(1/r_1,k_2)$.\\
 Secondly, we consider the region of $(k_2,\infty)$.
Because $P_I(u)\le 1$, $Q_B(u)>Q_B(k_2)=1$ and thus 
$R_B(u)=P_I(u)-Q_B(u)<0$ for all $u>k_2$,
so there is no fixed point in $(k_2,\infty)$ region.
Therefore, from the above two, 
we have proved that there is no fixed point in the region of 
$(1/r_1, \infty)$. $\Box$

In Theorem 4.1(a), we did not discuss the case when $k_2<1/r_2$. 
In Figure 2(a), we plot $R_B$ as function of $u$ for the cases that 
$M_b=0.5$ and $k_2=0.02$ (solid line), $k_2=0.04$ (dotted line),
$k_2=0.06$ (dashed line), $k_2=0.08$ (long dashed line).
All these values of $k_2$ satisfy  $k_2<1/r_2$. It is obvious that 
there is no root 
for $k_2=0.08$ (long dashed line) and two roots for others. 
From these numerical results, we know that there could be 
two fixed points or no fixed point in the $(0,1/r_2)$ region if $k_2<1/r_2$.

On the other hand, in Theorem 4.1(c)-(f), we did not discuss the case 
when $k_2/2<1/r_1<k_2$.
In Figure 2(b), we plot $R_B$ as function of $u$ for the cases that 
$M_b=0.5$ and $k_2=0.4$ (solid line), $k_2=0.5$ (dotted line),
$k_2=0.6$ (dashed line).
All these values of $k_2$ satisfy $k_2/2<1/r_1<k_2$.
It is obvious that there is no root 
for $k_2=0.4$ (solid line) and two roots for others. 
From these numerical results, we found that there could be 
two fixed points or no fixed point in 
in Region (III), i.e. $(1/r_1,\infty)$ if $k_2/2<1/r_1<k_2$.

Figure 2(c)-(d) are the numerical results for the solutions of equation 
(\ref{eq:II3}). That is, we find all possible fixed points 
$u$ for any given $k_2$ and $M_b$ with 
fixed values of $r_1$ and $r_2$. 

Figure 2(c) are the results on $k_2-u$ plane, where dashed lines are for
$M_b=0.1$, dotted lines are for $M_b=0.5$ and solid lines are for $M_b=1.0$.
From this figure, we know that there are three fixed points when $k_2$ 
is closer to $0$ or $1$, i.e. the left or right side of the figure.
However, there could be one fixed point only in the middle. 
Figure 2(d) are the results on $M_b-u$ plane, where dashed lines are for
$k_2=0.1$, dotted lines are for $k_2=0.5$ and solid lines are for $k_2=1.0$.

Checking the detail of these figures and Theorem 4.1, we found that
they are completely consistent with each other.

\subsection{Phase-Planes for Model B}
To study the phase planes, we need to know the eigenvalues of every fixed
point. The eigenvalues $\lambda$ corresponding to a fixed point 
$(u,v)$ satisfy the following equation:

$$\lambda^2-(-1+k_3 \frac{\pa B_1}{\pa u})=0,$$
\beq
{\rm where} \qquad \frac{\pa B_1}{\pa u}= \frac{\pa [\ln(P_I(u))/u]}{\pa u}
=-\frac{\ln P_I(u)}{u^2}+\frac{1}{uP_I(u)}\frac{\pa (P_I(u))}{\pa u} 
\label{eq:evb1}
\eeq
Hence 
\beq 
\lambda =\pm\sqrt{-1+k_3\frac{\pa B_1}{\pa u}}.
\eeq
  
We have two cases for the eigenvalues in the following:  

Case 1: If $-1+k_3\frac{\pa B_1}{\pa u}>0$, then $\lambda$ are real with 
opposite sign, so  the fixed point is a unstable saddle point.

Case 2: If $-1+k_3\frac{\pa B_1}{\pa u}<0$, then $\lambda$ are pure complex 
numbers with opposite sign, so the fixed point is a center point.

{\bf Theorem 4.2 }

If $k_2r_2<2$, then there is a unique fixed point 
$u_1\in (1/r_2, 1/r_1)$ and this fixed point is a center point.

{\bf Proof :} 

Since fixed point $(u,v)$ satisfy  equation (\ref{eq:II3}), from equation 
(\ref{eq:evb1}), we have  
\beqn
-1+k_3\frac{\pa B_1}{\pa u}&=&-1+k_3\left\{-\frac{\ln P_I(u)}{u^2}
+\frac{2}{u}\frac{(r_2-r_1)(1+r_1r_2u^2)}
{(1-r_2^2u^2)(1-r_1^2u^2)}\right\} \non \\
&=&-1+k_3\left\{-\frac{1}{k_3}+ \frac{k_2}{k_3u}+\frac{2}{u}
\frac{(r_2-r_1)(1+r_1r_2u^2)}{(1-r_2^2u^2)(1-r_1^2u^2)}\right\}\non \\
&=& -2+\frac{k_2}{u}+\frac{2k_3}{u}\frac{(r_2-r_1)(1+r_1r_2u^2)}
{(1-r_2^2u^2)(1-r_1^2u^2)} \non \\
&=& -2+\frac{k_2}{u}-\frac{k_2M_b(1+r_1r_2u^2)}
{u(r_2^2u^2-1)(1-r_1^2u^2)}.\label{eq:model_b_evtest}
\eeqn

Since $k_2/2$ is a global minimum of $Q_B(u)$ and $k_2/2 < 1/r_2$, 
we know that $Q'_B(u)>0$ for $u\in (1/r_2,1/r_1)$. 
Because $P'_I(u)<0$ and $Q'_B(u)>0$  for all $u\in (1/r_2,1/r_1)$, we have
$R'_B(u)=P'_I(u)-Q'_B(u)<0$ for all $u\in (1/r_2,1/r_1)$.
Moreover, it can be shown that 
$R_B((1/r_2)^{+})=\infty$ and $R_B(1/r_1)=-Q_B(1/r_1) <0$,
thus there is a unique fixed point $u_1\in (1/r_2,1/r_1)$ such that
 $R_B(u_1)=0$.

Because this fixed point $u_1\in (1/r_2,1/r_1)$, 
we have $$-2+\frac{k_2}{u_1}<-2+k_2r_2<0.$$
Since the third term of equation (\ref{eq:model_b_evtest}) is always negative, 
we have
$-1+k_3\frac{\pa B_1}{\pa u}\biggm|_{u=u_1}<0$. Therefore, 
this fixed point is a center point. $\Box$

%
%
%

The solution curves on the $u-v$ phase plane are shown in Figure 3(a)-(d),
where the vertical dotted lines are $u=1/r_2$ and $u=1/r_1$.
Figure 3(a) is the result when $k_2=0.05$, $M_b=0.5$, which corresponds
to the left region of Figure 2(c) (on the dotted line and $k_2$ is closer
to 0 ). Thus, there are three fixed points: one is larger than $1/r_2$
and others are less than $1/r_2$. 
Figure 3(b) is the result when $k_2=0.8$, $M_b=0.5$,  which corresponds
to the right region of Figure 2(c) (on the dotted line 
and $k_2$ is closer
to 1 ). Thus, there are three fixed points: one is less than $1/r_1$
and others are larger than $1/r_1$. 
Figure 3(c) is the result when $k_2=0.5$, $M_b=0.5$, which corresponds
to the middle region of Figure 2(c) (on the dotted line 
and about the region between 0.08 and 0.55) 
Thus, there is only one fixed point. 
Figure 3(d) is the result when $k_2=1.0$, $M_b=0.8$, which corresponds
to the right region of solid lines in Figure 2(d). Since $M_b$ is closer 
to 1, there are three fixed points: one is less than $1/r_1$
and others are larger than $1/r_1$. 
Therefore, the numbers of fixed points in Figure 3 are in fact completely 
consistent with  Figure 2. 

Figure 3(a), 3(c) and 3(b) 
give us a chance to see the transition of phase plane
along the dotted line in Figure 2(c) from left side to the right side.
In Figure 3(a), there are one center and one saddle point in the region 
of $u< 1/r_2$ and one center in the region of $1/r_2 < u < 1/r_1$.
However, in  Figure 3(c), the left two fixed points disappear, so there 
is only one fixed point, which is the center
in the region of $1/r_2 < u < 1/r_1$. Moreover, the shape of the solution 
curves  in the region of $u> 1/r_1$ already become different from the curves
close to a center. Finally, in Figure 3(b), there is still one center point
in the region of  $1/r_2 < u < 1/r_1$ but there are two new fixed points:
one saddle and one center in the region of $u> 1/ r_1$.
The locations of these bifurcation points, where transitions of phase plane
occur, can be easily seen from Figure 2(c).


These figures reconfirm the analytic results and also make the behavior of the 
solution curves clear.


\section{Model C}
In this model, we assume that the star-planet distance is between the inner
and outer radius of the belt, i.e $r_1<r<r_2$.
Thus, $f=f_s+f_{\alpha}+f_{b\epsilon}$ 
where $f_s$, $f_\alpha$ and $f_{b\epsilon}$ are defined in Section 2.

From equation (\ref{eq:1}), set $u=1/r$, we have 
\beq
\frac{d^2u}{d \theta^2}=-u+\frac{Gm^2}{l^2}-\frac{\alpha c}{lu}
\frac{du}{d\theta}-\frac{\pi Gcm^2}{l^2u}
\ln \left\{\biggm|\frac{(1+r_1u)(1+r_2u)}{(1-r_1u)(r_2u-1)}
\left(\frac{\epsilon^2u^2}{4-\epsilon^2u^2}\right)\biggm|\right\}. 
\label{eq:III1}
\eeq

To simplify the equation, we define 
\beq
P_\epsilon (u)\equiv\biggm|\frac{(1+r_1u)(1+r_2u)}{(1-r_1u)(r_2u-1)}
\left(\frac{\epsilon^2u^2}{4-\epsilon^2u^2}\right)\biggm|. 
\label{eq:p_epsilon}
\eeq

Hence, equation (\ref{eq:III1}) becomes
\beq
\frac{d^2u}{d\theta^2}+\frac{k_1}{u}\frac{du}{d\theta}=
-u+k_2-\frac{k_3}{u}\ln (P_\epsilon(u))  \label{eq:III2}
\eeq

where $k_1={(\alpha c)}/{l}$,
 $k_2={(Gm^2)}/{(l^2)}$ and $k_3={(c\pi Gm^2)}{(l^2)}$.
Therefore, we transform equation (\ref{eq:III2}) to the following problem

\beqn
 \frac{du}{d\theta}&=& v,  \non \\
\frac{dv}{d\theta}& =&-u+k_2-\frac{k_1}{u} v-\frac{k_3}{u}\ln(P_\epsilon(u)).
\label{eq:III3}
\eeqn

\subsection{Fixed Points for Model C}

The fixed points $(u,v)$ of problem (\ref{eq:III3}) satisfy the following 
equations

$$v=0$$
\beq
{\rm and}\qquad -u+k_2-\frac{k_1}{u}v-\frac{k_3}{u}\ln (P_\epsilon(u))=0.
\label{eq:III4}
\eeq

We define 
\beq
Q_c(u)\equiv\exp\left\{-\frac{u^2}{k_3}+\frac{k_2}{k_3}u\right\},
\eeq
and 
\beq
R_c(u)\equiv P_\epsilon(u)-Q_c(u),
\eeq
\no then the fixed points $(u,v)$ satisfy $R_c(u)=0$. In the 
following theorem, we discuss some properties about the fixed points.

{\bf Theorem 5.1}  

(a) There is at least one fixed point $u_1\in (0,1/r_2)$.  

(b) There is at least one fixed point
$u_{\epsilon\ast}\in (1/r_2,\bar{u})$, 
where $\bar{u}\in (1/r_2,1/r_1)$ such that $P'_\epsilon(\bar{u})=0$. Moreover, 
\beq
u_{\epsilon\ast}=\frac{1}{r_2}+\Od(\epsilon^2), {\,\,\rm so\,\,\,}
\lim_{\epsilon\to 0}u_{\epsilon\ast}=\frac{1}{r_2}.\label{eq:u_e_last}
\eeq

(c) There is at least one fixed point
$u^{\ast}_{\epsilon} \in (\bar{u},1/r_1)$, 
where $\bar{u}\in (1/r_2,1/r_1)$ such that $P'_\epsilon(\bar{u})=0$. Moreover, 
\beq
u_{\epsilon}^{\ast}=\frac{1}{r_1}+\Od(\epsilon^2), {\,\,\rm so\,\,\,}
\lim_{\epsilon\to 0}u^{\ast}_{\epsilon}=\frac{1}{r_1}.\label{eq:u_e_uast}
\eeq

(d) There are at least two fixed point $u_2,u_3\in (1/r_1,2/\epsilon)$.

{\bf Proof :}

(a) Since $R_c(0)=-Q_c(0)<0$ and $R_c((1/r_2)^{-})>0$, there is a root 
$u_1\in (0,1/r_2)$ such that $R_c(u_1)=0$. 
Thus, $u_1$ is a fixed point.$\Box$

(b) From equation (\ref{eq:p_epsilon}), if we consider $u\in (1/r_2,1/r_1)$, 
then we have 
\beq
P'_\epsilon(u)=\frac{2\epsilon^2 u}{(1-r_1u)(r_2u-1)(4-\epsilon^2u^2)}
\left\{-\frac{(r_1+r_2)(1-r_1r_2u^2)u}{(1-r_1u)(r_2u-1)}+
\frac{4(1+r_1u)(1+r_2u)}{(4-\epsilon^2u^2)}\right\}. \label{eq:dp_espilon}
\eeq

Since $P'_\epsilon((1/r_2)^{+})<0$, $P'_\epsilon((1/r_1)^{-})>0$ and 
$P''_\epsilon(u)>0$ for $u\in (1/r_2,1/r_1)$, 
there is a ${\bar{u}}\in (1/r_2,1/r_1)$ such that $P'_\epsilon({\bar{u}})=0$.

Since $P'_\epsilon({\bar{u}})=0$, from equation (\ref{eq:dp_espilon}), we have 
\beq
\frac{(r_1+r_2)(1-r_1r_2({\bar{u}})^2){\bar{u}}}{(1-r_1{\bar{u}})
(r_2{\bar{u}}-1)}=
\frac{4(1+r_1{\bar{u}})(1+r_2{\bar{u}})}{(4-\epsilon^2({\bar{u}})^2)}
\label{eq:dp_uast}
\eeq
Therefore, from equation (\ref{eq:dp_uast}), we have
\beqn
P_\epsilon({\bar{u}}) &=&
\frac{(1+r_1{\bar{u}})(1+r_2{\bar{u}})}{(1-r_1{\bar{u}})(r_2{\bar{u}}-1)}
\left(\frac{\epsilon^2({\bar{u}})^2}{4-\epsilon^2({\bar{u}})^2}\right)>0 
\non \\
&=& \frac{4\epsilon^2{\bar{u}}(1+r_1{\bar{u}})^2(1+r_2{\bar{u}})^2}
{(4-\epsilon^2({\bar{u}})^2)^2(r_1+r_2)(1-r_1r_2({\bar{u}})^2)}. 
\label{eq:p_e_uast}
\eeqn

From equation ({\ref{eq:p_e_uast}), we have 
$P_\epsilon({\bar{u}})={\rm{O}}(\epsilon^2)$.

Since $R_c\left((1/r_2)^{+}\right)>0$, and 
\beq
R_c(\bar{u})=P_\epsilon({\bar{u}})-Q_c(\bar{u})=
{\rm{O}}(\epsilon^2)-Q_c(\bar{u})<0, \label{eq:r_c_baru}
\eeq 
we have that there is at least one fixed point 
$u_{\epsilon\ast}\in (1/r_2, \bar{u})$ such that $R_c(u_{\epsilon\ast})=0$.

In the following, we prove that equation (\ref{eq:u_e_last}) holds. 
Since $R_c(u_{\epsilon\ast})=0$, we have 
$P_\epsilon(u_{\epsilon\ast})=Q_c(u_{\epsilon\ast})$, that is, 
\beq
\ln \left\{\frac{(1+r_1u_{\epsilon\ast})(1+r_2u_{\epsilon\ast})}
{(1-r_1u_{\epsilon\ast})(r_2u_{\epsilon\ast}-1)}
\left(\frac{\epsilon^2u^2_{\epsilon\ast}}{4-\epsilon^2u^2_{\epsilon\ast}}
\right)\right\}=-\frac{1}{k_3}u^2_{\epsilon\ast}
+\frac{k_2}{k_3}u_{\epsilon\ast}. \label{eq:pe_u_e_last} 
\eeq

We found the right hand side in equation (\ref{eq:pe_u_e_last}) is bounded 
for all $\epsilon >0$, since $u_{\epsilon\ast}\in (1/r_2,\bar{u})$.
We can rewrite the left hand side of 
equation (\ref{eq:pe_u_e_last}) to be
\beq
\ln \left\{\frac{1+r_1u_{\epsilon\ast}}{1-r_1u_{\epsilon\ast}}\right\}
+\ln \left\{\frac{1+r_2u_{\epsilon\ast}}{4-\epsilon^2u^2_{\epsilon\ast}}
\right\}
+\ln\left\{\frac{\epsilon^2u^2_{\epsilon\ast}}{r_2u_{\epsilon\ast}-1}\right\}.
\label{eq:pe_u_left}
\eeq
It is easy to show that the first term and the second term in equation 
(\ref{eq:pe_u_left}) are bounded for all $\epsilon >0$
since $u_{\epsilon\ast}\in (1/r_2,\bar{u})$. Therefore, the
third term in equation (\ref{eq:pe_u_left}) is bounded. That is, there exists
a $M>0$ such that 
\beq
-M<\ln\left\{\frac{\epsilon^2u^2_{\epsilon\ast}}
{r_2u_{\epsilon\ast}-1}\right\}<M.\label{eq:pe_left_bdd}
\eeq
Since $u_{\epsilon\ast}\in (1/r_2,\bar{u})$, from equation 
(\ref{eq:pe_left_bdd}), we have
\beq
1+\frac{\epsilon^2}{(1/\bar{u})^2\exp(-M)}<r_2u_{\epsilon\ast}<
1+\frac{\epsilon^2r_2^2}{\exp(M)}.
\eeq
Therefore, $u_{\epsilon\ast}=\frac{1}{r_2}+\Od(\epsilon^2)$ and 
$$\lim_{\epsilon\to 0}u_{\epsilon\ast}=\frac{1}{r_2}.\Box $$

(c) From equation (\ref{eq:r_c_baru}), we have $R_c(\bar{u})<0$ and 
$R_c((1/r_1)^{-})>0$, so there is at least one fixed point 
$u_{\epsilon}^{\ast}\in (\bar{u},1/r_1)$ such that 
$R_c(u_{\epsilon}^{\ast})=0$. 

For equation (\ref{eq:u_e_uast}), we have the same argument as the 
proof of (b).$\Box$

(d) From equation (\ref{eq:p_epsilon}), if we consider 
$u\in (1/r_1,2/\epsilon)$ then 
we have 
\beq
P'_\epsilon(u)=\frac{2\epsilon^2 u}{(r_1u-1)(r_2u-1)(4-\epsilon^2u^2)}
\left\{-\frac{(r_1+r_2)(r_1r_2u^2)u-1}{(r_1u-1)(r_2u-1)}+
\frac{4(1+r_1u)(1+r_2u)}{(4-\epsilon^2u^2)}\right\}. \label{eq:dp1_espilon}
\eeq

 Since $P'_\epsilon(\{1/r_1\}^{+})<0$, $P'_\epsilon(\{2/\epsilon\}^{-})>0$ and 
$P''_\epsilon(u)>0$ for $u\in (1/r_2,1/r_1)$, 
there is a ${\hat{u}}\in (1/r_1,2/\epsilon)$ such that 
$P'_\epsilon({\hat{u}})=0$
and $P_\epsilon({\hat{u}})$ is a minimum value for $u\in (1/r_1,2/\epsilon)$.

Since $P'_\epsilon({\hat{u}})=0$, from equation (\ref{eq:p_epsilon}), we have 
\beq
\frac{(r_1+r_2)(r_1r_2({\hat{u}})^2-1){\hat{u}}}{(r_1{\hat{u}}-1)
(r_2{\hat{u}}-1)}=
\frac{4(1+r_1{\hat{u}})(1+r_2{\hat{u}})}{(4-\epsilon^2({\hat{u}})^2)}
\label{eq:dp_hatu}
\eeq
Therefore, from equation (\ref{eq:dp_hatu}), we have
\beqn
P_\epsilon({\hat{u}}) &=&
\frac{(1+r_1{\hat{u}})(1+r_2{\hat{u}})}{(r_1{\hat{u}}-1)(r_2{\hat{u}}-1)}
\left(\frac{\epsilon^2({\hat{u}})^2}{4-\epsilon^2({\hat{u}})^2}\right)>0 
\non \\
&=& \frac{4\epsilon^2{\hat{u}}(1+r_1{\hat{u}})^2(1+r_2{\hat{u}})^2}
{(4-\epsilon^2({\hat{u}})^2)^2(r_1+r_2)(r_1r_2({\hat{u}})^2-1)}, 
\label{eq:p_e_hatu}
\eeqn

and we found the minimum value 
$P_\epsilon({\hat{u}})={\rm{O}}(\epsilon^2)$. Therefore, 
$R_c(\hat{u})=P_\epsilon(\hat{u})-Q(\hat{u})<0$. 
Since $R_c((1/r_1)^{+})>0$, $R_c((2/\epsilon)^{-})>0$ and $R_c(\hat{u})<0$,
there are at least two root $u_2\in (1/r_1,\hat{u})$ and 
$u_3\in (\hat{u},2/\epsilon)$ such that $R_c(u_2)=R_c(u_3)=0$. $\Box$

\subsection{Phase-Planes for Model C}

Next, we find the eigenvalues $\lambda$ satisfy the following equation
\beq
\lambda^2+\frac{\alpha c}{lu} \lambda-(-1-k_3 \frac{\pa C_1}{\pa u})=0,
\label{eq:lambda_c}
\eeq

\beq
{\rm where \,\, } C_1=\frac{\ln(P_\epsilon(u))}{u}. 
\eeq
Hence,
\beq 
\frac{\pa C_1}{\pa u}= \frac{\pa 
[\ln(P_\epsilon(u))/u]}{\pa u}
=-\frac{\ln P_\epsilon(u)}{u^2}+\frac{1}{uP_\epsilon(u)}
\frac{\pa (P_\epsilon(u))}{\pa u}. 
\label{eq:dC_1}
\eeq
%
%
%

Thus, from equation (\ref{eq:lambda_c}), we have
\beq
\lambda=\frac{\alpha c}{2lu}\left\{-1\pm\sqrt{1-4
\left(\frac{\alpha u}{lc}\right)^2\left(1+k_3\frac{\pa C_1}{\pa u}\right)}
\right\}.
\eeq

Let $\Delta=1-4\left(1+k_3\frac{\pa C_1}{\pa u}\right)
\left(\frac{lu}{\alpha c}\right)^2$, then we have three possible cases :

Case 1: If $\Delta<0$, then two eigenvalues 
$\lambda_{1,2}=-a\pm b {\rm i}$ for $a,b>0$, so 
the fixed point is a stable spiral point.

Case 2: If $0<\Delta<1$, then two eigenvalues are negative, so
the fixed point is a stable node.

Case 3: If $\Delta>1$, then one eigenvalue is positive and the other is 
negative, so the fixed point is a saddle point.


{\bf Theorem 5.2}  

(a) The fixed point, $u_{\epsilon\ast}=\frac{1}{r_2}  
+\Od(\epsilon^2)$ is a saddle point.  

(b) The fixed point, $u_{\epsilon}^{\ast}=\frac{1}{r_1}
+\Od(\epsilon^2)$ is a stable spiral point.  

{\bf Proof :} 

Since fixed point $(u,v)$ satisfy equation (\ref{eq:III4}), from equation 
(\ref{eq:dC_1}), we have 
\beqn
&&1+k_3\frac{\pa C_1}{\pa u} \non \\
&&=1+k_3\left\{
-\frac{\ln P_\epsilon(u)}{u^2}-\frac{2(r_1+r_2)(1-r_1r_2u^2)}
{u(1-r_1^2u^2)(r_2^2u^2-1)}+\frac{8}{u^2(4-\epsilon^2u^2)}\right\} \non \\
&&=2-\frac{k_2}{u}-\frac{2k_3(r_1+r_2)(1-r_1r_2u^2)}
{u(1-r_1^2u^2)(r_2^2u^2-1)}+\frac{8k_3}{u^2(4-\epsilon^2u^2)}. 
\eeqn

In order to know which case it should be for different fixed point, 
we further calculate: 
\beqn
\Delta &= &1-4\left(1+k_3\frac{\pa C_1}{\pa u}\right)
\left(\frac{lu}{\alpha c}\right)^2, \non \\
&=& 1-4\left( 2-\frac{k_2}{u}-\frac{2k_3(r_1+r_2)(1-r_1r_2u^2)}
{u(1-r_1^2u^2)(r_2^2u^2-1)}+\frac{8k_3}{u^2(4-\epsilon^2u^2)}\right)
\left(\frac{lu}{\alpha c}\right)^2, \non \\
&=& 1-8\left(\frac{lu}{\alpha c}\right)^2+
\left(\frac{4k_2l^2u}{\alpha^2 c^2}\right)+
\frac{8k_3l^2 u(r_1+r_2)(1-r_1r_2u^2)}
{\alpha^2c^2(1-r_1^2u^2)(r_2^2u^2-1)}
-\frac{32 k_3l^2}{\alpha^2 c^2(4-\epsilon^2u^2)}. \label{eq:Delta} 
\eeqn

(a) We consider the fixed point, $u_{\epsilon\ast}=\frac{1}{r_2}
+\Od(\epsilon^2)$. Since the fourth term of equation (\ref{eq:Delta}) is 
positive with  magnitude $\Od(1/\epsilon^2)$ and other terms are bounded
, thus $\Delta>1$ and the fixed point 
$u_{\epsilon\ast}=\frac{1}{r_2}+\Od(\epsilon^2)$ is a saddle point. 
$\Box$

(b) We consider the fixed point, $u_{\epsilon}^{\ast}=\frac{1}{r_1}
+\Od(\epsilon^2)$. Since the fourth term of equation (\ref{eq:Delta}) is 
negative with magnitude $\Od(1/\epsilon^2)$ and other terms are bounded
, thus $\Delta<0$ and the fixed point 
$u_{\epsilon}^{\ast}=\frac{1}{r_2}+\Od(\epsilon^2)$ is a stable spiral point. 
$\Box$

The solution curves on the $u-v$ phase plane are shown in Figure 4(a)-(d),
where we set $k_2=0.1$, $M_b=0.5$ and $\epsilon=10^{-3}$. 
In Figure 4(a),  there are two vertical dotted lines, the left one 
is $u=1/r_2$ and the right one is $u=1/r_1$, which divide the $u-v$
plane into three regions. 
Figure 4(b) is the detail for the region near the line of $u=1/r_2$ 
in Figure 4(a), which make it possible to see the fine structures.
We can see that there is one center and one saddle point near the dotted
line $u=1/r_2$.
Figure 4(c) is the detail for the region near the line of $u=1/r_1$ 
in Figure 4(a). 
We can see that there is one spiral point and one saddle point near the dotted
line $u=1/r_1$.
Figure 4(d) is a closer-looking of the spiral point in the region of 
$u>1/r_1$.


These figures reconfirm the analytic results and also make the behavior of the 
solution curve clear.

\section{Conclusions}

We have studied the properties of solution curves on the phase plane
of dynamical systems of planet-belt interaction by the standard
fixed-point analysis. 


The system is divided into three
regions: (a) the region between the central star and the inner edge of 
the belt, (b) the region out of the outer edge of the belt,
(c) the region between the inner edge and the outer edge of 
the belt. The dynamics 
in three regions are governed by three different models and the planet 
moves around these regions. 

We analytically prove some properties of the system and also
show some bifurcation diagrams of fixed points for different cases:
the number and properties of fixed points changed with the
values of parameters $k_2$ and $M_b$ in general.

Our analytical and numerical results show that, in most cases, 
the locations of fixed points depend on the values of these 
parameters and these fixed points are either 
center (structurally stable) or saddle (unstable).  

On the other hand,  we found that 
there are two special fixed points:
one is on the inner and another is on the outer edges of the belt. 
The one on the inner edge is a stable focus (asymptotically  stable) and 
the one on the outer edge is a saddle (unstable).

This interesting result is consistent with the 
observational picture of Asteroid Belt between the Mars and Jupiter: 
the Mars is moving stablely close to the inner edge but the Jupiter is
quite far from the outer edge.

\clearpage

\begin{figure}[tbhp]
\epsfysize 7.0in \epsffile{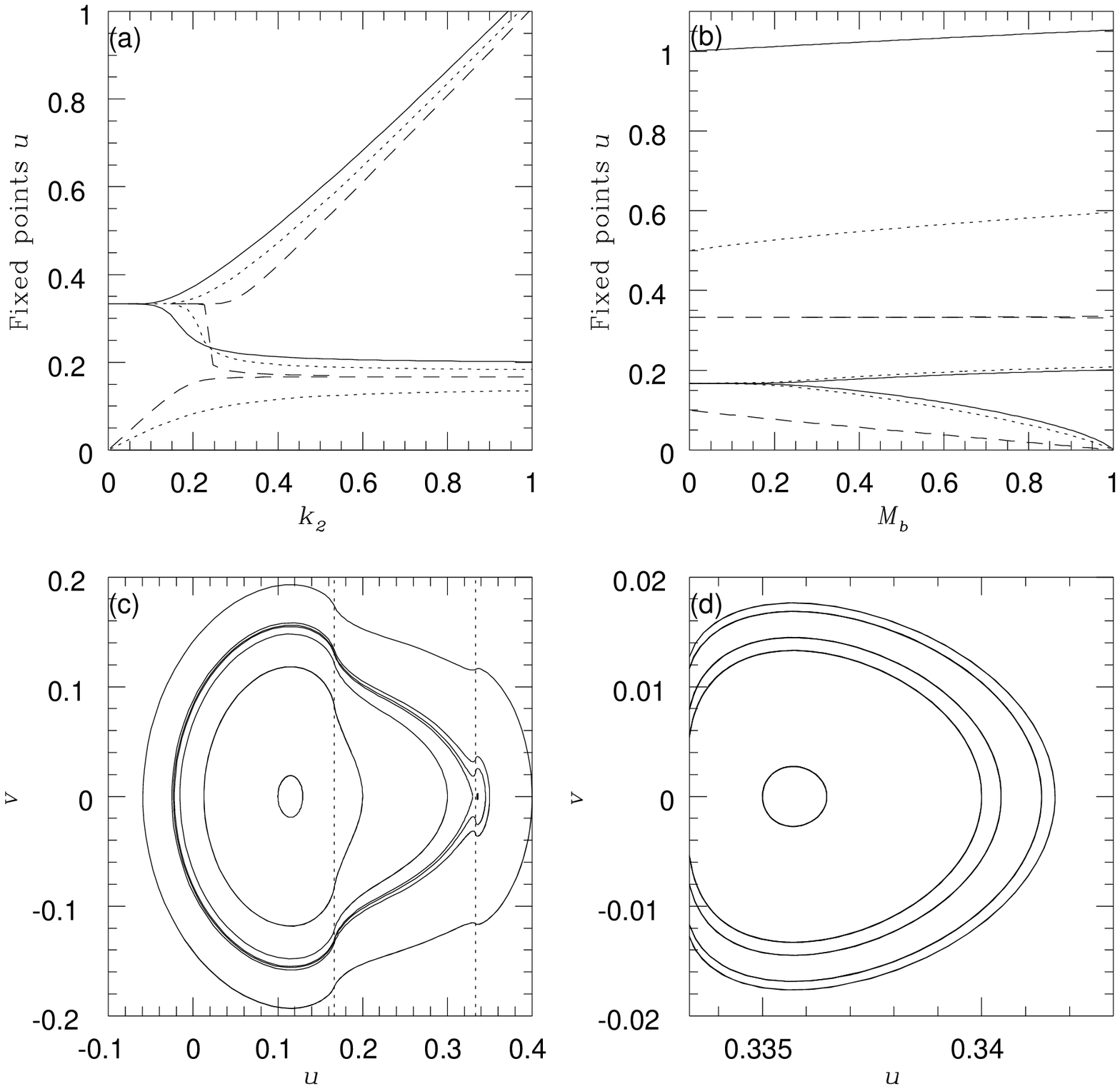}
\caption{
Bifurcation diagrams of fixed points and phase planes
for Model A with $r_1=3$, $r_2=6$.
(a) The bifurcation diagram of fixed points on $k_2-u$ plane, 
where dashed lines are for $M_b=0.1$, 
dotted lines are for $M_b=0.5$ and solid lines are for $M_b=1.0$.
 (b) The bifurcation diagram of fixed points on $M_b-u$ plane, 
where dashed lines are for $k_2=0.1$, 
dotted lines are for $k_2=0.5$ and solid lines are for $k_2=1.0$.
 (c) The solution curves on the $u-v$ plane, 
where we set $k_2=0.2$, $M_b=0.3$. There are two vertical dotted lines, 
the left one is $u=1/r_2$ and the right one is $u=1/r_1$.
(d) The detail near the rightest fixed points of (c).
}
\end{figure}

\clearpage

\begin{figure}[tbhp]
\epsfysize 7.0in \epsffile{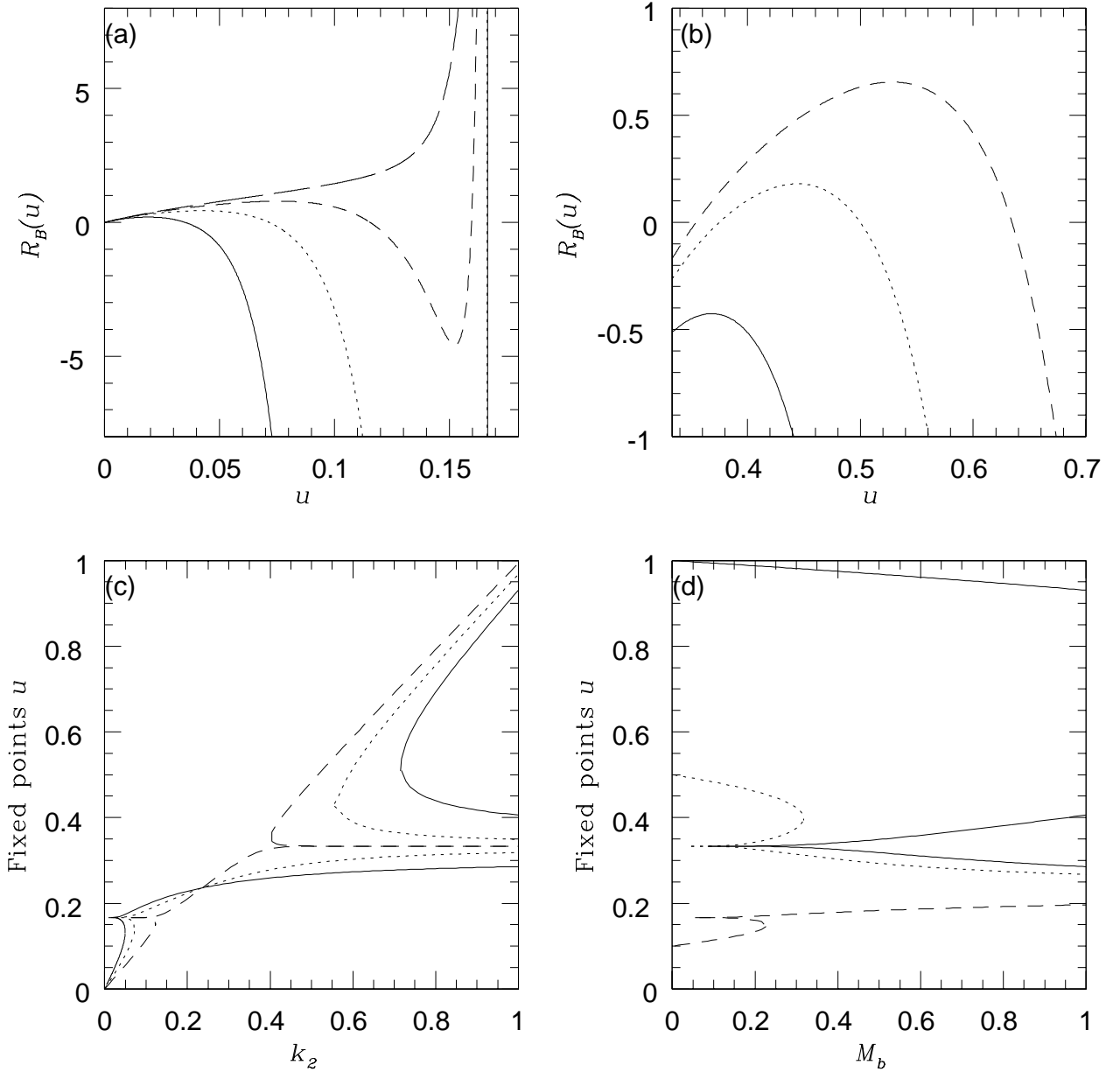}
\caption{
Bifurcation diagrams of fixed points for Model B with $r_1=3$, $r_2=6$.
(a)  $R_B$ as a function of $u$ for the cases that 
$M_b=0.5$ and $k_2=0.02$ (solid line), $k_2=0.04$ (dotted line),
$k_2=0.06$ (dashed line), $k_2=0.08$ (long dashed line).
(b) $R_B$ as a function of $u$ for the cases that 
$M_b=0.5$ and $k_2=0.4$ (solid line), $k_2=0.5$ (dotted line),
$k_2=0.6$ (dashed line).
(c)  The fixed points on $k_2-u$ plane, 
where dashed lines are for $M_b=0.1$, 
dotted lines are for $M_b=0.5$ and solid lines are for $M_b=1.0$.
(d) The fixed points on $M_b-u$ plane, 
where dashed lines are for $k_2=0.1$, 
dotted lines are for $k_2=0.5$ and solid lines are for $k_2=1.0$.
}
\end{figure}

\clearpage

\begin{figure}[tbhp]
\epsfysize 7.0in \epsffile{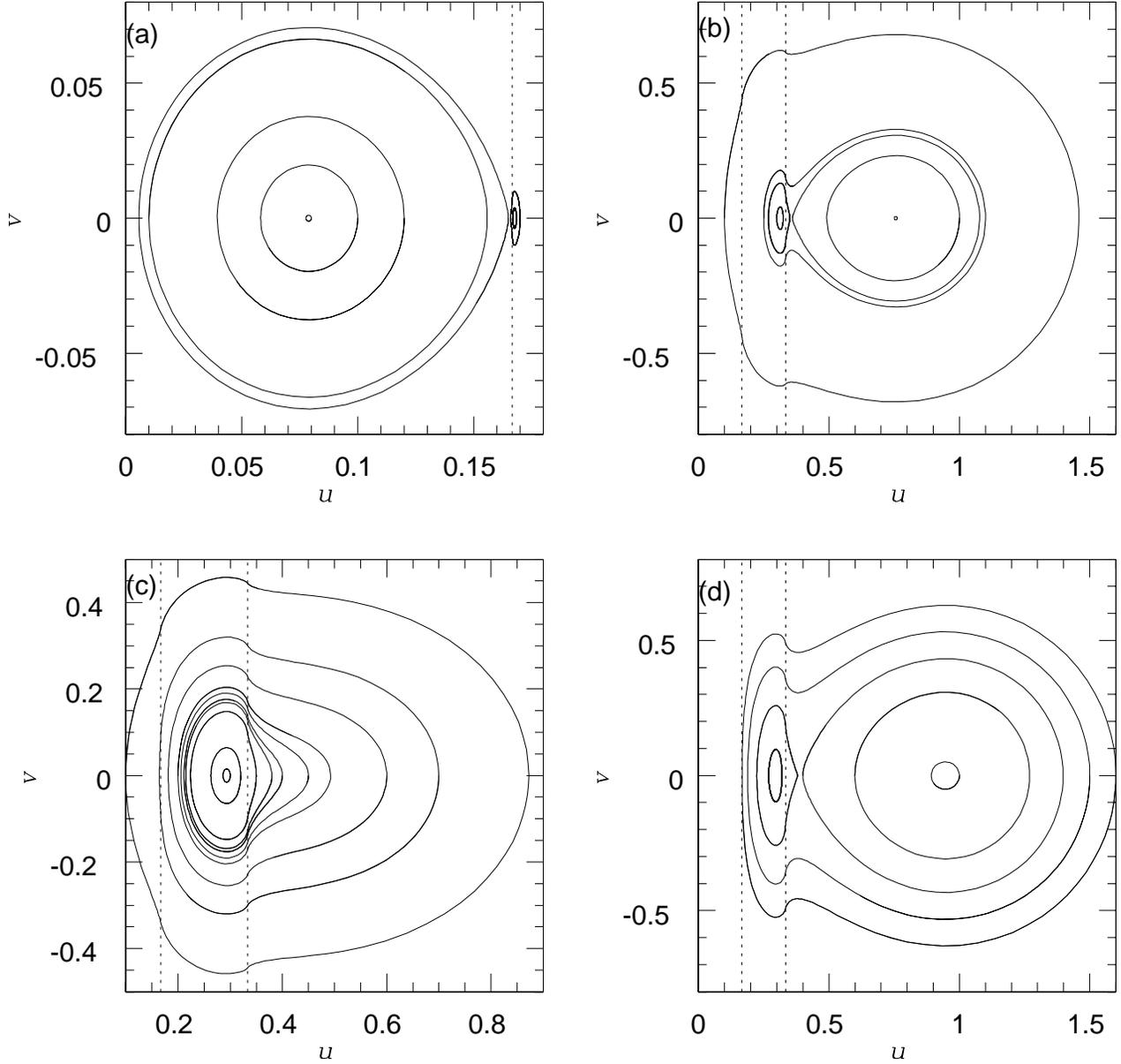}
\caption{
The phase planes for Model B with $r_1=3$, $r_2=6$.
(a) The solution curves on the $u-v$ plane when 
    $k_2=0.05$, $M_b=0.5$. The vertical dotted line is $u=1/r_2$. 
 (b) The solution curves on the $u-v$ plane when 
    $k_2=0.8$, $M_b=0.5$. The left vertical dotted line is $u=1/r_2$  
     and  the right vertical dotted line is $u=1/r_1$. 
 (c)  The solution curves on the $u-v$  plane when 
      $k_2=0.5$, $M_b=0.5$. The left vertical dotted line is $u=1/r_2$  
     and  the right vertical dotted line is $u=1/r_1$. 
(d) The solution curves on the $u-v$ plane when 
      $k_2=1.0$, $M_b=0.8$. The left vertical dotted line is $u=1/r_2$  
     and  the right vertical dotted line is $u=1/r_1$.
}
\end{figure}

\clearpage

\begin{figure}[tbhp]
\epsfysize 7.0in \epsffile{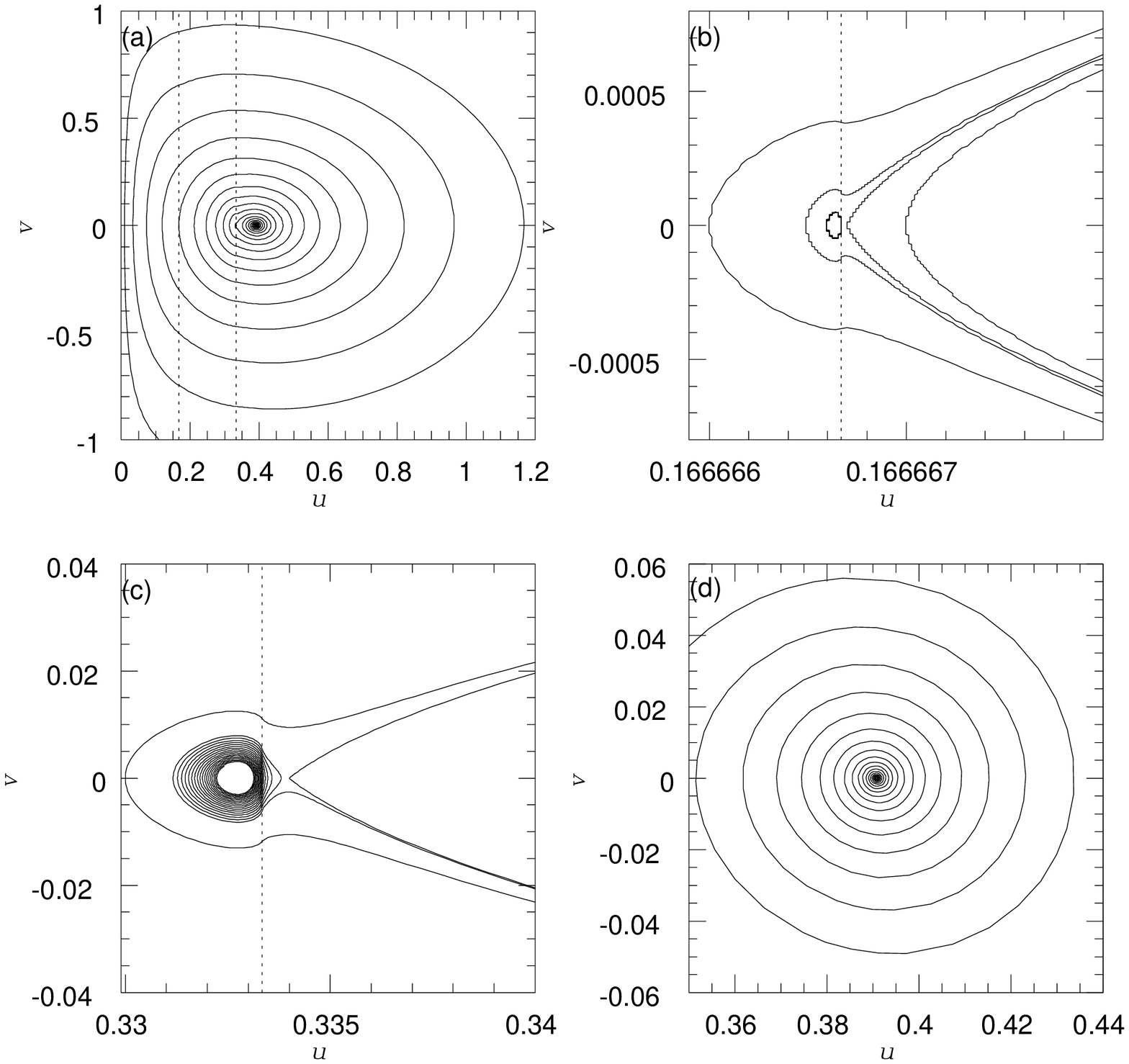}
\caption{
The phase planes for Model C when $r_1=3$, $r_2=6$, $k_2=0.1$, $M_b=0.5$ 
and $\epsilon=10^{-3}$. (a) The $u-v$ phase plane, the left vertical line 
is $u=1/r_2$ and the right one is $u=1/r_1$.
 (b) The detail for the region near the line of $u=1/r_2$ 
      in (a).
(c) The detail for the region near the line of $u=1/r_1$ 
in (a). 
(d) A closer-looking of the focus point in the region of $u>1/r_1$.
}
\end{figure}


\begin{thebibliography}{21}

\bibitem{ChanChungQi} Chan, H.S.Y. \& Chung, K.W. and Dongwen, Qi 
[2001] ``Some 
bifurcation diagrams for limit cycles of Quadratic differential systems'',
Int. J. Bifurcation and Chaos {\bf 11}, 197-206.


\bibitem{CHS} Clausen, S. \& Helgesen, G. and Skjeltorp, A.T. [1998]
``Braid description of few body dynamics'',
Int. J. Bifurcation and Chaos {\bf 7}, 1383-1397.


\bibitem{Gol}  Goldstein, H. [1980]  ``Classical Mechanics'', Addison-Wesley 
Publishing Company 


\bibitem{JiangIp} Jiang, I.-G. \& Ip, W.-H. [2001] ``The planetary system
of upsilon Andromedae'', Astronomy \&  Astrophysics {\bf 367}, 943-948. 

\bibitem{KIM} Kaulakys, B. \& Ivanauskas, F. and Meskauskas, T. [1999]
``Synchronization of chaotic systems driven by identical noise'',
Int. J. Bifurcation and Chaos {\bf 9}, 533-539.

\bibitem{Thom} Thommes, E.W. \& Duncan, M.J. and Levison, H.F. [1999]
``The formation of Uranus and Neptune in the Jupiter-Saturn region 
of the Solar System'', Nature {\bf 402}, 635-638

\bibitem{YehJiang} Yeh, L.-C. \& Jiang, I.-G. [2001] ``Orbital
Evolution of Scattered Planets'', The Astrophysical Journal {\bf 561},
364-371.



 

\end{thebibliography}
\end{document}